\documentstyle[prl,aps,epsf,floats]{revtex}

\newcommand{\mbg}[1]{\mbox{$\boldmath #1$}}
\begin{document}
\draft
\twocolumn[\hsize\textwidth\columnwidth\hsize\csname @twocolumnfalse\endcsname
\title {Quasielastic magnetic scattering of neutrons 
at the systems with heavy fermions} 

\author{Andrei S. Mishchenko}
 
\address{RRC "Kurchatov Institute" 123182 Moscow, Russia}

\date{\today}

\maketitle

\begin{abstract}
The theory of the quasielastic magnetic scattering of neutrons
at the spin liquid with RVB (resonance valence bonds)
correlations is presented. Calculations demonstrate that the
dependence of the scattering cross section on the energy
transfer reproduces experimental shape of quasielastic peak in
heavy fermion systems. It is shown that Fermi statistics of
the spin liquid elementary excitations leads to the oscillations
of the quasielastic scattering total cross section as a function
of the momentum transfer.
\end{abstract}

\pacs{Accepted to JETP Lett., Vol.\ 68}
\vskip2pc]
\narrowtext

1. Since it was suggested \cite{Bas87,Can89,KiKiM97} to treat
the state with heavy fermions (HF) as the spin liquid (SL) with
RVB type of spin-spin correlations the validity of RVB
conception is still the subject of considerable discussions.
Although conception of RVB type SL appeared to be fruitful for
quantitative description of the thermodynamic \cite{cjp96} and
low energy spectral
\cite{KaKiM97} properties of HF systems there is no unambiguous
proof of existence of RVB correlations. 
The basic stumbling-block on the way to the SL identification is
the lack of decisive experiment which can give possibility to
accept or reject RVB conception.
\par
According to the RVB scenario the system of localized spins
transforms at temperatures close to Kondo temperature $T_K$ into
the half filled band of magnetic excitations with the bandwidth
$T^* \sim T_K$ \cite{KiKiM97,Grel92,Tan93}.
The basic fingerprint of the RVB correlations, which distinguishes
SL state from the system of localized spins, is the {\it Fermi
statistics} of the elementary excitations.
Therefore, the study of consequences of the different
statistical distributions which characterize SL and localized
spin regime is a proper way to reveal some properties which are
peculiar only to the highly correlated RVB state.
\par
In the present paper I suggest the theory of the quasielastic magnetic 
scattering of neutrons at the RVB type SL and show that change 
of the Boltzmann statistics of localized spins to the Fermi statistics 
of the SL excitations results in the oscillatory behavior of the 
total quasielastic cross section as a function of momentum transfer. 
\par
2. The basic model describing the HF state is the Anderson
lattice Hamiltonian of the f-ions which are hybridized with the
conduction electrons.
The Coqblin - Schrieffer canonical transformation
\cite{Coqb69,Corn72} eliminates the hybridization term and the
lowest crystal field doublet $\sigma = \pm$ can be treated in
terms of exchange interaction $I_{\bf{mm}'}^{\sigma\sigma'}$
between the spin states of different sites $\bf{m} =
\bf{l},\mbg{\xi}$ ($\bf{l}$ is the number of the elementary cell,
$\mbg{\xi}$ is the basis vector)
\begin{equation}
H_{ex} = 
\sum_{\bf{mm}'}^{\bf{m} \ne \bf{m}'} 
\sum_{\sigma\sigma'}
I_{\bf{mm}'}^{\sigma\sigma'}
f^{\dagger}_{\bf{m}\sigma} f_{\bf{m}\sigma'}  
f^{\dagger}_{\bf{m}'\sigma'} f_{\bf{m}'\sigma}  
\label{ham}
\end{equation}
Here $f_{\bf{m}\sigma}^{\dagger}$ ($f_{\bf{m}\sigma}$) are
creation (annihilation) Fermi operators of the spin state
$\sigma$ at the site $\bf{m}$ which are subject to the constraint
condition $\sum_{\sigma} f^{\dagger}_{\bf{m}\sigma}
f_{\bf{m}\sigma} = 1$.
\par 
In the mean field approximation the RVB state at low temperatures is 
characterized by nonzero intersite averages 
$\langle f^{\dagger}_{\bf{m}\sigma} f_{\bf{m}'\sigma} \rangle$ and  
effective Hamiltonian of the SL low temperature states takes the form
\begin{equation}
H_{ex}^{ef} = 
\sum_{\bf{mm}'}^{\bf{m} \ne \bf{m}'} 
\sum_{\sigma}
A_{\bf{mm}'}
f^{\dagger}_{\bf{m}\sigma} f_{\bf{m}'\sigma},
\label{efha}
\end{equation}
which formally resembles tight binding approximation for band spectrum.
The constants $A_{\bf{mm}'}$ are determined by the values of
exchange integrals $I_{\bf{mm}'}^{\sigma\sigma'}$ and anomalous
averages \cite{KaKiM97}.
The chemical potential $\mu$ in the mean field approximation can
be deduced from the global constraint for $N$-ion lattice
$N^{-1} \sum_{\bf{m}}\sum_{\sigma}
\langle f^{\dagger}_{\bf{m}\sigma} f_{\bf{m}\sigma} \rangle  = 1$.
Therefore, the RVB SL in the mean field approximation can be
considered as a half filled band of Fermi quasiparticles with
the width $\sim T^*$.
The eigenstates $\mid \lambda \rangle$ of the Hamiltonian (\ref{efha})
can be expressed in terms of eigenvectors $\Xi_{\xi\sigma}^{\lambda}$ 
in the form of superposition \cite{KaKiM97}
\begin{equation}
\mid \lambda \rangle = \sum_{\bf{l}\xi\sigma}
\exp(i\bf{l}\bf{k_{\lambda}}) \Xi_{\xi\sigma}^{\lambda} 
\varphi(\bf{r}-\bf{l}-\mbg{\xi}) \mid \sigma \rangle,
\label{eig}
\end{equation}
(Here $\varphi(\bf{r}-\bf{l}-\mbg{\xi})$ is the spatial wave
function of localized f-electron; $\mid \sigma \rangle$ is the
spin component of the wave function and $\bf{k_{\lambda}}$ is
the wave vector of the state $\mid \lambda \rangle$.)
The occupation numbers of the states $\mid \lambda \rangle$ with
the energies $\varepsilon_{\lambda}$ at temperature $T$ are 
described by Fermi distribution $n_{\lambda} =
\{1+\exp[(\varepsilon_{\lambda}-\mu)/T] \}^{-1}$.
\par
3. The standard expression for the scattering function \cite{ML}
per one magnetic ion with the momentum transfer $\mbg{\kappa}$
and energy transfer $\hbar\omega$ can be expressed in terms of the
matrix elements of operator $\hat{\bf{Q}} = \sum_{\bf{l}\xi}
\exp \{i\mbg{\kappa}(\bf{l}+\mbg{\xi}) \} \hat{\bf{J}}$ and Cartesian  
components of unit vectors 
$\hat{\mbg{\kappa}}_{\alpha} = \mbg{\kappa}_{\alpha} / \mid \mbg{\kappa}\mid$
\begin{eqnarray}
&S(\mbg{\kappa},\hbar\omega) = 
b^2 N^{-1}
\sum\limits_{\alpha\beta}^{xyz}
(\delta_{\alpha\beta} - \hat{\mbg{\kappa}}_{\alpha} \hat{\mbg{\kappa}}_{\beta})
\times & \nonumber \\
&\sum\limits_{\lambda\lambda'}
\Omega_{\lambda,\lambda'}
\langle \lambda \mid \hat{Q}_{\alpha}^{\dagger} \mid \lambda' \rangle 
\langle \lambda' \mid \hat{Q}_{\beta} \mid \lambda \rangle 
\delta(\hbar\omega-\varepsilon_{\lambda'}+\varepsilon_{\lambda})&
\label{ssta}
\end{eqnarray}
($b$ is the magnetic scattering length \cite{ML}; $\hat{\bf{J}}$ is the  
operator of total magnetic moment). In case of the scattering at the 
localized crystal field states the statistic function 
$\Omega_{\lambda,\lambda'}$ is connected with the Boltzmann
distribution and depends only on the occupation numbers of the
initial states $\lambda$.  On the other hand in case of
Fermi statistics of SL excitations the occupation of the final
state is also important and statistic function takes the form
\begin{equation}
\Omega_{\lambda,\lambda'} = n_{\lambda} (1-n_{\lambda'}).
\end{equation}
Inserting the eigenstates (\ref{eig}) into expression (\ref{ssta}) and
neglecting the intesite matrix elements of operator $\hat{\bf{J}}$ between
highly localized $f$-electron states one gets the following expression for the 
scattering function
\begin{eqnarray}
&S_{sl}(\mbg{\kappa},\hbar\omega) = 
B(\kappa)
\sum\limits_{\alpha\beta}^{xyz}
(\delta_{\alpha\beta} - \hat{\mbg{\kappa}}_{\alpha} \hat{\mbg{\kappa}}_{\beta})
\times & \nonumber \\
&\sum\limits_{\lambda\lambda'}
\Omega_{\lambda,\lambda'}
\Theta^{\alpha\beta}_{\lambda\lambda'}(\mbg{\kappa})
\delta(\hbar\omega-\varepsilon_{\lambda'}+\varepsilon_{\lambda}),&
\label{spf}
\end{eqnarray}
where
\begin{equation}
\Theta^{\alpha\beta}_{\lambda\lambda'}(\mbg{\kappa}) = 
I_{\alpha}^{\lambda\lambda'}(\mbg{\kappa}) 
\left( I_{\beta}^{\lambda\lambda'}(\mbg{\kappa}) \right)^*,
\label{the}
\end{equation}
and
\begin{equation}
I_{\alpha}^{\lambda\lambda'}(\mbg{\kappa}) =
\sum\limits_{\mbg{\xi}}\exp(-i\mbg{\kappa}\mbg{\xi}) 
\sum\limits_{\sigma\sigma'} 
\left( \Xi_{\xi\sigma}^{\lambda} \right)^*
\Xi_{\xi\sigma'}^{\lambda'}
\langle \sigma \mid \hat{J}_{\alpha} \mid \sigma' \rangle.
\label{i}
\end{equation}
Here $B(\kappa) = b^2 \left( g_J F(\kappa)/2 \right)^2$ where
$g_J$ is the Lande factor and $F(\kappa)$ is the formfactor of
the magnetic ion.
\par
4. First of all it have to be noted that dependence of the
scattering function (\ref{spf}) on the energy transfer
$\hbar\omega$ and temperature qualitatively describes the experimentally 
situation with the quasielastic peaks in HF systems \cite{FuLo85}.
The quasielastic peakshape in HF systems is treated usually by use of
the phenomenological expression \cite{FuLo85}
\begin{equation}
S_{ph}(\hbar\omega,T) = 
\frac{1}{1-\exp(-\hbar\omega/T)}
\frac{\hbar\omega}{\Gamma(T)^2+(\hbar\omega)^2} .
\label{phes}
\end{equation}
Characteristic feature of HF systems is nonzero width $\Gamma(T)$ at zero 
temperature.
\par
For quantitative estimation let us consider the simple case of
symmetric spinon band with the constant density of states ${\cal
D}(\varepsilon) = 1$ for $\varepsilon \in [-T^*,T^*]$ and ${\cal
D}(\varepsilon) = 0$ out of the defined range.
The chemical potential of this band is defined by condition of
half filling and equal to zero.
After integrating of the scattering function (\ref{spf}) over
the angles (which corresponds to the scattering at a
polycrystal) one gets the following expression
\begin{eqnarray}
&S_{sl}(\kappa, \hbar\omega) = &
\nonumber \\
&R(\kappa) \int\limits^{T^*}_{-T^*} 
d \varepsilon {\cal D}(\varepsilon + \hbar\omega) 
n(\varepsilon) (1-n(\varepsilon+\hbar\omega)).&
\label{som}
\end{eqnarray}
The momentum dependence is determined by the factor $R(\kappa)$
and will be considered in the next section. 
Now we concentrate on the dependence of the 
scattering function on the energy transfer which for the defined above 
density of states ${\cal D}(\varepsilon)$ has the following form 
$$
S_{sl}(\kappa, \hbar\omega) =
\frac{R(\kappa)}{2} 
T \exp \left( \frac{\hbar\omega}{2T} \right) 
\sinh^{-1} \left( \frac{\mid \hbar\omega \mid }{2T} \right)
\times
$$
\begin{equation}
\ln \left\{
\frac{1+\cosh\left[ T^* / T \right]}
{1+\cosh\left[(T^*- \mid \hbar\omega \mid ) / T \right]}
\right\}
\label{sen}
\end{equation}
\par
\begin{figure}[t]
\begin{center}
\leavevmode
\hbox{
\epsfxsize=8.4cm
\epsffile{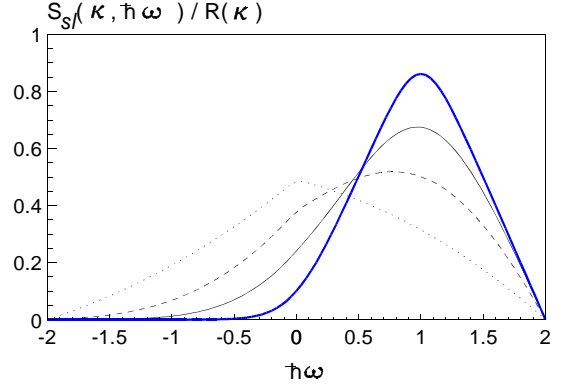}
}
\end{center}
\vspace{-0.5cm}
\caption{ 
The energy dependence of the scattering function of the spin liquid 
with the bandwidth $T^*=2$ for $T=0.1$ (thick solid line), $T=0.25$ 
(thin solid line), $T=0.5$ (dashed line) and  $T=2$ (dotted line).
}
\label{fig1}
\end{figure}

As may be seen from Fig.\ \ref{fig1} the energy and temperature dependence 
of expression (\ref{sen}) is in qualitative agreement with experimental 
measurements of quasielastic response in HF systems. 
Moreover, the energy dependence (\ref{sen}) is similar for 
$\hbar\omega < T^*$ to the phenomenological function (\ref{phes}).
It is instructive that generally accepted for the experimental
data treatment fit of the formula (\ref{phes}) to the scattering
function (\ref{sen}) gives $\Gamma(T=0) \sim T^*$.
It have to be noted that breaks of the energy dependence at $\hbar\omega=0$
and restriction of $S_{sl}(\kappa, \hbar\omega)$ by the energy range 
$[-2T^*,2T^*]$ can be avoided by use of less singular density of states
and involving the damping of SL excitations, respectively.
\par
5. The energy dependence of the quasielastic scattering at the SL 
can not be used as a decisive argument in favor of RVB model applicability 
for HF systems because there are lot of models which give the same result 
(e.g. \cite{Bic87,KisMis97}). However, much more bright consequence of the 
RVB model, which can serve as decisive test, is the dependence of the 
scattering function on the momentum transfer.
\par
To demonstrate this consequence of the Fermi statistic let us
consider two-sublattice system in which intersublattice
interaction dominates, i.e.
$A_{\bf{l}\mbg{\xi},\bf{l}'\mbg{\xi}'} = \delta_{\bf{l}\bf{l}'}
T^*/2$ for $\mbg{\xi} \ne \mbg{\xi}'$ and negligible for
$\mbg{\xi} = \mbg{\xi}'$.
(Numeric calculations show that these approximations, although
simplify the analytic estimations, do not influence the
character of presented effects.)
In the framework of this model the four-fold degeneracy of the
states in each elementary cell is reduced and the set of
eigenstates consists of two two-fold degenerate eigenstates with
the energies $\pm T^*/2$. 
The chemical potential is $\mu=0$.  
Let us assume for definitness sake that the basis vector
$\mbg{\xi}$ is parallel to $z$-axis and the distance between the
magnetic ions is $d$.
\par
To calculate the momentum dependence it is convenient to introduce the 
function
\begin{equation}
S_{sl}^{qe}(\mbg{\kappa} ) = 
\int^{T^*}_{-T^*} d(\hbar\omega) 
S_{sl}(\mbg{\kappa},\hbar\omega),
\label{to}
\end{equation}
which corresponds to the integral value of the cross section with respect to 
the energy transfer. This function represents 
the total quasielastic scattering on the lowest crystal field level.  
\par
In the final expression for the SL scattering function
\begin{equation}
S_{sl}^{qe}(\mbg{\kappa}) = 
\left( \frac{B(\kappa)}{2} \right) \sum_{\alpha\beta} 
(\delta_{\alpha\beta} - \hat{\mbg{\kappa}}_{\alpha} \hat{\mbg{\kappa}}_{\beta})
T^{\alpha\beta} D(\mbg{\kappa},T),
\label{toc}
\end{equation}
where 
\begin{equation}
T^{\alpha\beta} = \sum_{\sigma\sigma'}
\langle \sigma \mid \hat{J}_{\alpha} \mid \sigma' \rangle
\langle \sigma' \mid \hat{J}_{\beta} \mid \sigma \rangle,
\label{ani}
\end{equation}
the {\it Fermi statistics results in the additional factor} 
\begin{equation}
D(\mbg{\kappa},T) = 
\frac{1}{2}
\left\{
1-\tanh^2\left( \frac{T^*}{T} \right)  \cos (\mbg{\kappa} \bf{d})
\right\},
\label{osc}
\end{equation}
which is {\it equal to unity in case of Boltzmann distribution}.
The factor $D(\mbg{\kappa},T)$ leads at low temperatures to the 
oscillation of the scattering cross section with the period
$\kappa_p = 2 \pi /( d \cos [\widehat{\mbg{\kappa}\bf{d}}])$. 
Important property for experimental check is the dependence of the 
oscillations period on the momentum transfer direction. 
\par
The scattering function for polycrystalline samples $S_{sl}^{tot}(\kappa)$
can be calculated by integration of expression (\ref{toc}) over the 
angles. To demonstrate the consequences of Fermi statistic I separated out
into the factor $S^{tot}(\kappa)$ the standard scattering function for 
Boltzmann statistics which only source of the momentum transfer dependence 
is the square of formfactor $F(\kappa)$.  
It is seen from the following expression 
$$
S_{sl}^{tot}(\kappa) = S^{tot}(\kappa) \times
$$
\begin{equation}
\frac{1}{2} \left\{ 1 -  \tanh^2\left( \frac{T^*}{T} \right) \left[
\left(
1+\frac{\eta}{2}
\right)
\frac{\sin(\kappa d)}{\kappa d}
- \frac{3}{2} \eta \Phi(\kappa d) 
\right]
\right\}
\label{tot}
\end{equation}
that integration does not lead to the total suppression of oscillations.  
Here 
\begin{equation}
\eta = 
\frac{T^{xx}+T^{yy}-2T^{xx}}{T^{xx}+T^{yy}+T^{xx}} 
\label{peta}
\end{equation}
is the measure of the one site anisotropy of scattering.
The character of the oscillations is determined by the factor 
$\sin(\kappa d)/(\kappa d)$ and function
\begin{equation}
\Phi(\kappa d) =  \frac{\sin(\kappa d)-\kappa d\cos(\kappa d)}{(\kappa d)^{3}}.
\label{phi}
\end{equation}
For large enough momentum transfer $\kappa d \gg 1$ the function
$\Phi(\kappa d)$ is much less than factor $\sin(\kappa d)/(\kappa d)$. 
\begin{figure}[t]
\begin{center}
\leavevmode
\hbox{
\epsfxsize=8.4cm
\epsffile{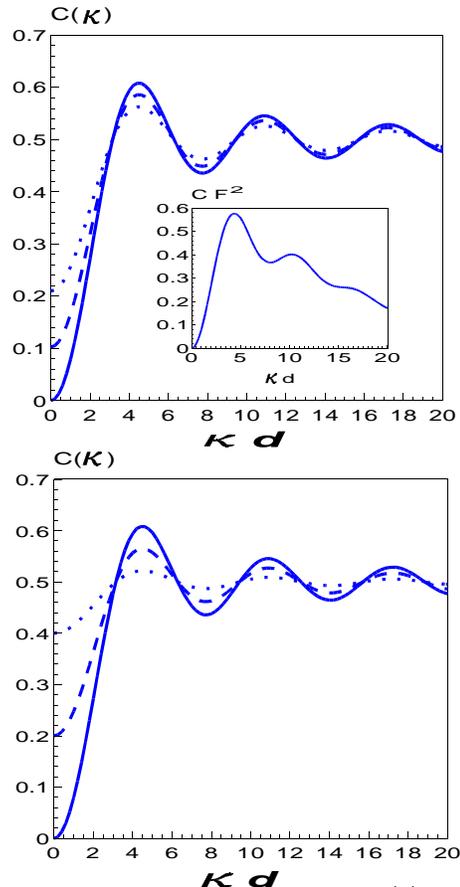}
}
\end{center}
\vspace{-0.5cm}
\caption{ 
Dependence of the ratio 
$'(\kappa)= S_{sl}^{tot}(\kappa)/S^{tot}(\kappa)$ 
on the momentum transfer module $\kappa$ for SL bandwidth $T^*=1$ and
$\eta=0$. 
Upper panel: in the perfect lattice 
$T=0$ (solid line), $T=0.7$ (dashed line) and $T=1.0$ (dotted line).
Lower panel: at zero temperature for $\Lambda=1$ (solid line), 
$\Lambda=0.6$ (dashed line) and $\Lambda=0.2$ (dotted line). Insert: 
Scattering function taking into account the formfactor of 
Ce$^{3+}$ ion for the interionic distance $d=4 \AA$.
}
\label{fig2}
\end{figure}
It is seen from the upper panel of Fig.\ \ref{fig2} that the
temperature does not significantly suppress the oscillations,
at least in the mean field approximation. Even at temperature 
$\sim T^*$, which is enough to destroy SL state, the oscillations 
amplitude is still noticeable.
\par
One should note that the period of driven by the Fermi statistic
oscillations is determined by the distance between interacting
magnetic ions $d$ and have nothing to do with the peculiarities
of SL Fermi surface. The characteristic distance between the
magnetic ions in the HF compounds is $d \sim 4 \AA$. Therefore,
in spite scattering function decrease $S_{sl}^{tot}(\kappa) 
\sim F^2(\kappa)$ it is possible to observe
several oscillation periods (see insert in the upper panel of
Fig.\ \ref{fig2}).
\par
6. However, the influence of the lattice imperfection on the suppression of
the oscillations is much more significant.
In the simplest case, when coherency is not destroyed by the
absence of one of the magnetic ions, the lattice imperfection results in 
different environments of basis ions.
The change of environment leads to the change of the wave
function of the crystal field states \cite{Mis97}.
Different wave functions of basis ions lead, 
in turn, to inequality of matrix elements
$\langle \sigma \mid \hat{J}^{\alpha} \mid \sigma' \rangle_1 \ne
\langle \sigma \mid \hat{J}^{\alpha} \mid \sigma' \rangle_2$ 
for sites 1 and 2.
Although expression for scattering function is derived for general case I 
present less cumbersome formula for the isotropic one-ion scattering case
($x$, $y$ and $z$ directions are equivalent and Cartesian indices are omitted
in the following). Nonequivalence of the basis ions can be characterized by
the quantity
\begin{equation}
U = 
\frac{
\sum\limits_{\sigma\sigma'} 
\left|
\langle \sigma \mid \hat{J} \mid \sigma' \rangle_1 -
\langle \sigma \mid \hat{J} \mid \sigma' \rangle_2 
\right|^2
}
{
\sum\limits_{\sigma\sigma'} 
\left|
\langle \sigma \mid \hat{J} \mid \sigma' \rangle_1 +
\langle \sigma \mid \hat{J} \mid \sigma' \rangle_2 
\right|^2
}
.
\label{ne}
\end{equation}
Within this simple model the influence of the lattice imperfection on the 
scattering function 
\begin{equation}
S_{sl}^{tot}(\kappa) = S^{tot}(\kappa) \frac{1}{2}
\left\{
1- \Lambda \tanh^2\left( \frac{T^*}{T} \right)
\frac{\sin(\kappa d)}{\kappa d}
\right\}
\label{totim}
\end{equation}
is determined by the factor 
\begin{equation}
\Lambda = \frac{1-U}{1+U}, 
\label{lambda}
\end{equation}
which is equal to unity 
in the perfect lattice and tends to zero for the total lost of coherency.
The lower panel of Fig.\ \ref{fig2} demonstrates the influence of the 
lattice imperfection on the suppression of the oscillations.
\par
7. In conclusion, it have to be noted that dependence of the
spin liquid neutron scattering cross section on the energy
transfer can not be distinguished from the scattering on the
relaxing spin.
However, the oscillatory dependence of the cross section on the momentum 
transfer is unique for spin liquid of RVB type. 
Besides, in case of soft enough crystal field splitting
$\Delta_{CF} \sim T^*$ one have to include into the Hamiltonian 
(\ref{ham}) the terms which are nondiagonal with respect to crystal
field levels \cite{KaKiM97}. This terms lead to the oscillations of the 
inelastic scattering intensity. The oscillations of the inelastic
scattering intensity were experimentally observed \cite{Sato95} 
in Kondo -  semimetal CeNiSn and quantitatively explained in the 
framework of spin liquid conception \cite{KaKiM97}. 
However, since many characteristics of the periodic systems demonstrate 
the oscillatory behavior in neutron scattering response 
(e.g. the peak width oscillations in paramagnets \cite{DeGe58,NevRai96}),
it is important to find the property which is connected with RVB correlations
unambiguously. 
\par
As it is shown in the present paper, the oscillations of the
total quasielastic scattering cross section is direct
consequence of the Fermi statistic of RVB elementary
excitations. Therefore, an experimental observation of the
predicted effect might serve as notable argument in favor of the
validity of spin liquid conception for the description of heavy
fermion systems.
\par
I am deeply grateful to P.\ A.\ Alekseev, K.\ A.\ Kikoin and 
V.\ N.\ Lazukov for critical discussions. This work was supported by
Russian Fund for Fundamental Research (project No.\ 98-02-16730).

\end{document}